\documentclass[preprint,showpacs,preprintnumbers,amsmath,amssymb]{revtex4}

\usepackage{graphicx}
\usepackage{dcolumn}
\usepackage{bm}

\begin{document}

\preprint{APS/123-QED}

\title{Finite-temperature magnetism of Fe$_x$Pd$_{1-x}$ and Co$_x$Pt$_{1-x}$ alloys}

\author{S.~Polesya$^a$,\; S.~Mankovsky$^a$,\; O.~Sipr$^b$,\; W.~Meindl$^c$,\; C.~Strunk$^c$,\; H.~Ebert$^a$}

\affiliation{
$^a$ Department of Chem.\ and Biochem./Phys.\ Chem., LMU Munich,\\
Butenandtstrasse 11, D-81377 Munich, Germany  \\
$^b$ Institute of Physics AS~CR, v.~v.~i., Cukrovarnicka~10,
162~53~Praha, Czech Republic \\
$^c$ University Regensburg, D-93040 Regensburg,  Germany
}

\date{\today}

\begin{abstract}
The finite-temperature magnetic properties of Fe$_x$Pd$_{1-x}$ and
Co$_x$Pt$_{1-x}$ alloys have been investigated. It is shown that
the temperature-dependent magnetic behaviour of alloys, 
composed of originally magnetic and non-magnetic elements,
cannot be described properly unless the coupling between
magnetic moments at magnetic  atoms (Fe,Co) mediated through the
interactions with induced magnetic  
moments of non-magnetic atoms  (Pd,Pt) is included.
A scheme for the calculation of the Curie temperature ($T_C$) for
this type of systems is presented which is based on the extended
Heisenberg Hamiltonian with the appropriate exchange parameters $J_{ij}$
obtained from 
{\em ab-initio} electronic structure calculations. Within the present study the KKR Green's
function method has been used to calculate the  $J_{ij}$ parameters. A
comparison of the obtained Curie temperatures for  
Fe$_x$Pd$_{1-x}$ and Co$_x$Pt$_{1-x}$ alloys with experimental data
shows rather good agreement.
\end{abstract}


\maketitle

\section{Introduction}

Whether a magnetic material is technologically useful or not
depends on its properties at finite temperatures.  However, the {\em
  ab-initio} treatment of finite-temperature magnetism remains a challenge
despite ongoing progress in this field during the last decades. 

In this context, itinerant-electron
3d-transition metals and their alloys  receive  particular interest.
For these systems, finite-temperature magnetic properties cannot be
 described successfully neither within the collective-electron Stoner
model nor within the local-moment model based on the Heisenberg
Hamiltonian (for an overview see Refs. \cite{Kueb00,Mohn03}). 
 The Stoner model, which
treats the transition to the paramagnetic state as vanishing of the
local magnetic moments, accounting thereby only for the longitudinal
spin fluctuations, grossly overestimates the Curie temperature
$T_C$.  A bigger success has been achieved using the Heisenberg model,
which accounts for temperature-induced transverse spin fluctuations
and characterizes the paramagnetic state by
orientational disorder within the system of localized magnetic moments.
However, a magnitude of the moments 
is assumed to be unchanged upon fluctuations. In some recent studies the
Heisenberg model approach has been combined with {\em ab-initio}
band-structure calculations, that allow to evaluate the exchange
coupling parameters from first principles \cite{LKAG87}. In this way,
trends of the Curie or N\'{e}el temperature with composition can be
quantitatively described for many systems, in 
particular, for transition metal monoxides \cite{FDE09}, dilute
magnetic semiconductors \cite{FSK04} or  transition metals
\cite{TKDB03}.

Despite the rather satisfying results obtained within this combined
approach for several systems, the description of finite-temperature
magnetism of transition
metals and alloys still suffers from many problems owing to the
restrictions of the Heisenberg model. 
For some itinerant-electron systems, e.g. Ni, the thermally-induced
longitudinal spin fluctuations play a crucial role in describing
properly the temperature-dependent magnetisation and obtaining the
correct value for the critical temperature.  
 A phenomenological theory of finite temperature magnetism,
which accounts for both types of fluctuations on the same footing, was
developed in the past \cite{Hub79,Mor85,Has79,KMP77}.  This theory was
used in combination with {\em ab-initio} electronic structure
calculations to describe temperature dependent magnetic properties of
Fe, Co and Ni \cite{UK96,RJ97,RKMJ07}.  In this way a much
better agreement with experiment, as compared to calculations based on
the Heisenberg model, was obtained.  In particular, proper accounting for
longitudinal fluctuations results in the vanishing of local magnetic
moments on Ni atoms above $T_C$ \cite{RKMJ07}, in agreement with
experiment.

Other interesting itinerant-electron systems in this context  are alloys or 
compounds composed of originally magnetic and non-magnetic elements. 
Such systems exhibit the so called covalent magnetism
\cite{WZM+81,MS93}, where magnetisation of the 'non-magnetic' atoms is
governed by the spontaneously magnetised atoms via the strong
spin-dependent hybridisation of their electronic states. 
The  Fe$_x$Pd$_{1-x}$ and Co$_x$Pt$_{1-x}$ alloys considered in the
present work  belong to this type of systems.
To describe the temperature-dependent magnetism of such systems on the
basis of Heisenberg model, one obviously has to account properly for the
behavior of the Pd/Pt sub-lattices. Only a few {\em ab-initio} studies of
finite-temperature magnetism of systems of this kind have been done so
far. Similarly to the work mentioned above \cite{UK96,RJ97,RKMJ07} these studies were based on a
generalisation of the classical  
Heisenberg Hamiltonian in one or another way to account for the
different character of magnetism on different types of atoms.  Mryasov
et al. \cite{MNG05} have investigated the compound FePt and showed that the anomalous
temperature-dependence of its magneto-crystalline anisotropy energy
(MAE) is due to the induced Pt magnetic moments.  In another study
of these authors \cite{Mry05}, a
crucial role of the magnetic moment induced on Rh was demonstrated for the
stabilisation of the ferromagnetic state of FeRh and for the control of
the antiferromagnet-ferromagnet phase transition.  Lezaic et
al. \cite{LME+06} emphasized the need to account
for longitudinal fluctuations of magnetic moments induced on Ni atoms
for a proper description of the temperature-dependence of the
spin-polarisation at the Fermi energy $E_{F}$\ in the half-metallic
ferromagnet NiMnSb.  Sandratskii et al. \cite{SSS07} investigated several ways
for accounting for the induced magnetic moments within the spin-spiral
approach used for the calculation of the exchange
coupling parameters in NiMnSb and MnAs. Using these results Sandratskii
et al. \cite{SSS07} have studied finite-temperature magnetic properties of NiMnSb;
their findings are consistent with the findings of Lezaic et al \cite{LME+06}.

In this work we introduce  an {\em
    ab-initio} method  to describe finite-temperature
magnetism of systems with spontaneous and induced magnetic moments. The 
method is based on an extension of the Heisenberg Hamiltonian by adding
a term which describes the induced magnetic moments within the
linear response formalism.  Our approach relies on a combination of
{\em ab-initio} band-structure calculations with Monte Carlo (MC)
simulations based on the extended Heisenberg model.
To test this approach, we  investigate 
finite-temperature magnetic properties of Pd-rich  Fe$_x$Pd$_{1-x}$
alloys, with 
Fe concentrations up to 20 at.\%, as well as of ordered and disordered
Co$_{3}$Pt, CoPt and CoPt$_{3}$ alloys that are 
interesting both for fundamental reasons and for possible use in
industrial applications because of their high magnetic anisotropy
\cite{CHZ+00,MLR+08}. Theoretical investigations of finite-temperature
magnetism of these systems failed so far to reproduce experimental
results with satisfactory accuracy \cite{KGS99}.
We demonstrate in the present work that a combination of  {\em ab-initio} band-structure
calculations with the Monte Carlo (MC) 
simulations based on the extended Heisenberg model gives satisfying
agreement between theoretical and experimental values of critical temperatures.
We found that despite their small magnitudes, the moments induced on
non-magnetic atoms (Pd, Pt) have an important influence on
finite-temperature magnetic order.

\section{Experimental detailes}

The Fe$_x$Pd$_{1-x}$ films were thermally evaporated onto oxidized
silicon substrates from separate effusion cells for Pd and Fe in an
ultra high vacuum system (base pressure $5 \dot 10^{-11}$ mbar). The
film thicknesses were between 15 and 20 nm. The deposition rate of the
two components could be controlled independently, resulting in an
accuracy of the Fe concentration of $1\%$. Auger spectroscopy on a
sample with nominally $7\%$ Fe content provided an independent value of
$7.8\%$ Fe for this film. After deposition, the films were patterned
into a six-terminal Hall-bar geometry. Measurements of the anomalous
Hall effect at temperatures between 2K and 300 K provided the
magnetization $M(T)$, from which the Curie temperatures of the films
were deduced.

\section{Theoretical approach}
\subsection{Ground-state calculations}

Within the present work, spin-polarised  electronic structure
calculations for the ground-state have  been performed using the
multiple scattering KKR (Korringa-Kohn-Rostoker) Green's function method
\cite{Ebe00} in the scalar-relativistic
approximation. The local spin density approximation (LSDA) for
 density functional theory was used with the 
parametrisation for the exchange-correlation potential due to Vosko,
Wilk, and Nusair  \cite{VWN80}. 
The potential was treated within the
atomic sphere approximation (ASA) with the radii of the spheres around Fe/Co
and Pd/Pt sites chosen by requiring the ratios of the corresponding
volumes to be the same as for the pure elements.  For the angular momentum
expansion of the Green's function a cutoff of
$l_{max} = 3$ was applied.  For substitutionally
disordered alloys, the self-consistent coherent potential approximation (CPA)
method was employed. A geometry optimisation was performed, i.e., the
lattice constants of the alloys have been obtained by minimisation of
the total energy.

\subsection{Extended Heisenberg Hamiltonian}

The finite temperature properties of the investigated systems were
studied by Monte Carlo simulations based on the Heisenberg model with the
underlying Hamiltonian given by

\begin{eqnarray}
\label{Heisenberg}
H_{ex} &=& -\sum_{ij}\tilde{J}^{M-M}_{ij}\vec{M}_i\vec{M}_j 
 -\sum_{ij}\tilde{J}^{M-m}_{ij}\vec{M}_i\vec{m}_j  \nonumber \\
&& -\sum_{ij}\tilde{J}^{m-m}_{ij}\vec{m}_i\vec{m}_j\;.
\end{eqnarray}
%

Here the classical Hamiltonian was generalised to allow an application
to itinerant-electron systems consisting of magnetic and non-magnetic
atoms having magnetic moments $M_i$ and $m_i$, respectively, connected
with corresponding exchange coupling parameters. The dependence of the
induced magnetic moments $m_i$ on a specific magnetic configuration is
treated via linear response formalism (for details see the Appendix). 

We suppose that the induced magnetic moments on Pd or Pt atoms are
governed only by the magnetic moments of the Fe or Co atoms in
Fe$_x$Pd$_{1-x}$ and Co$_x$Pt$_{1-x}$ alloys, respectively, arranged in
the first neighbor shell around the non-magnetic atom, so that

\begin{eqnarray}
\label{A3}
\vec{m}_i &=& \sum_{j\in M} X^{m-M}_{ij}\vec{M}_j =  X_i^{m-M} \sum_{j\in M} \vec{M}_j \;.
\end{eqnarray}
The notation $\sum_{j\in M}$\ means that the sum includes only such
terms where $j$\ is a site with an inducing magnetic moment.

In this case (see Appendix)  
the susceptibility  $X^{m-M}_{ij}$ can be approximated by the value found for the ground
state of a system with well defined collinear spontaneous (Fe, Co) and induced
(Pd, Pt) magnetic moments:

\vspace{-0.5cm}
\begin{eqnarray}
\label{A4}
 X_i^{m-M} &=& \frac{m_i}{\sum_{j\in M}{M}_j}\;.
\end{eqnarray}

The exchange coupling parameters $\tilde{J}_{ij}$ between atoms $i$\ and
$j$\ in the Eq. (\ref{Heisenberg}) were obtained via the formula of
Lichtenstein et al. \cite{LKAG87}:

 \begin{equation} 
\label{eq_jij}
J_{ij} \: = \: -\frac{1}{4\pi} \, \mbox{Im} \int^{E_F} \mathrm{d} E\; \mathrm{Tr} \:
 (t^{-1}_{i\uparrow} -t^{-1}_{i\downarrow}) \,
\tau^{ij}_{\uparrow} \, (t^{-1}_{j\uparrow} -t^{-1}_{j\downarrow})
\, \tau^{ji}_{\downarrow} \; .
\end{equation}

with the relation

\begin{equation} 
\label{eq_jij1}
\tilde{J}_{ij} = \frac{J_{ij}}{|\vec{m}_i||\vec{m}_j|} \; .
\end{equation}

In Eq.~(\ref{eq_jij}), $t_{i\sigma}$ and $\tau^{ij}_{\sigma}$ are the spin ($\sigma$) and site  ($i, j$) dependent single-site and scattering path operator matrices occurring within the KKR formalism \cite{Wei90}.

\subsection{Evaluation of $T_{C}$}
\label{meth-Ac}

The Curie temperature $T_C$ was evaluated with the Monte Carlo (MC) method
\cite{Bin97} using the standard Metropolis importance sampling
algorithm \cite{LB00}, on the basis of the model Hamiltonian in Eq. (\ref{Heisenberg}). The
number of atoms in the MC unit cell for different concentrations was
taken between 1728 and 4000 and $T_C$ was 
determined from the peak position of the temperature-dependent susceptibility. The Fe/Co
magnetic moments were treated during MC simulation as localised 
and changed only their orientation. On the other hand, the magnetic
moments on Pd/Pt atoms could change their absolute value as well as the
orientation in accordance with the changed magnetic configuration around
these atoms. Eq. (\ref{A3}) implies that the magnetic moments on Pd/Pt are proportional to
the vector sum of magnetic moments at neighboring Fe/Co atoms with only
nearest neighbors taken into account. This means that each MC step
consists of 1) change of the orientation of a magnetic moment on the Fe/Co
atoms, 2) search for all nearest neighbour Pd/Pt atoms and calculation of the
orientation and absolute value of their moments using the susceptibilities
$X^{m-M}$ via Eq. (\ref{A4}).
The change of the energy of the total system is due to both effects. 
For disordered alloys the resulting $T_C$ values in addition were averaged over up to 20 different configurations.

Obviously, the approach described above, 
accounts for the contribution of spin polarised
'non-magnetic' atoms to the exchange interactions between the magnetic
atoms. As this contribution is 
temperature-dependent, it allows a corresponding description of the
temperature dependent magnetisation. In particular, it 
accounts for longitudinal spin fluctuations occurring on the
'non-magnetic' sub-lattice. 
Below it will be shown that this rather simple scheme
gives rather good 
agreement with experimental data for the systems  under consideration.

\section{Results for  Fe$_x$Pd$_{1-x}$ alloys}

\subsection{{\em Ab-initio} calculations}
\label{}

\begin{figure}[h]
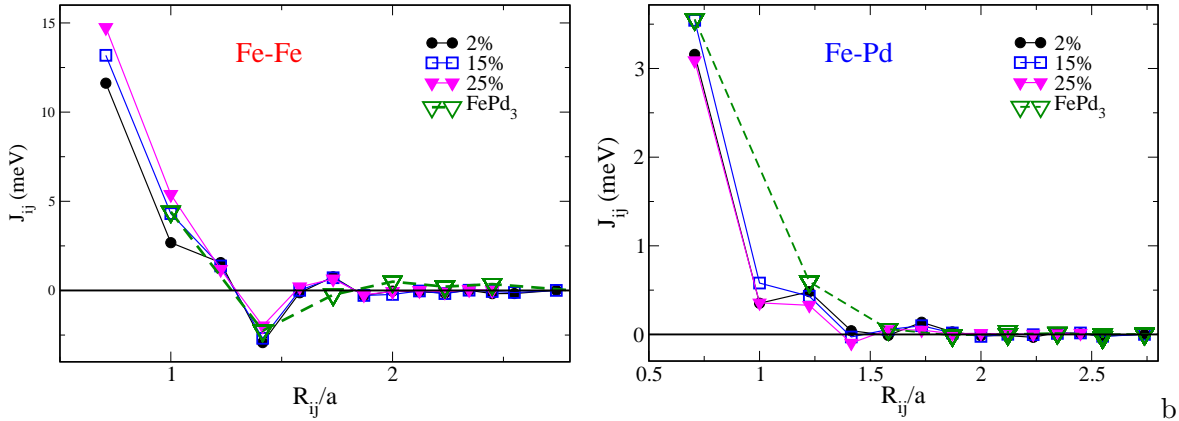

\includegraphics[width=7.5cm,angle=0]{Jij_fefe_diso.eps}a
\includegraphics[width=7.5cm,angle=0]{Jij_fepd_diso.eps}b  

\caption{\label{Fig1} Exchange coupling parameters $J^{Fe-Fe}$ (a) and
  $J^{Fe-Pd}$ (b)  for Fe$_{x}$Pd$_{1-x}$ alloys at different
  concentrations. The dashed lines represent results for
  ordered FePd$_3$ (note that in this case for the Fe atoms
  the nearest neighbour interaction $J^{Fe-Fe}$ is absent because there
  are only Pd nearest neighbours). }
\end{figure}

The scheme for calculation of temperature dependent magnetic
properties, described above, was used to investigate disordered  Fe$_x$Pd$_{1-x}$ alloys
with Fe concentration up to 20 at.\%.
The exchange coupling parameters $J^{Fe-Fe}_{ij}$ and  $J^{Fe-Pd}_{ij}$,
shown in Fig. \ref{Fig1}, have a similar dependency on the distance
$R_{ij}$ for all investgated alloys. 
In the Pd reach limit (Fe concentration $x < 0.2$)
 the exchange coupling parameters $J^{Fe-Fe}_{ij}$ corresponding to the
 average distance  between  magnetic atoms is rather small and does
 not allow to create long-range magnetic order in the system, as was
 demonstrated by corresponding restricted MC simulations.


\vspace{2cm}


\vspace{5cm}
\begin{figure}[h]
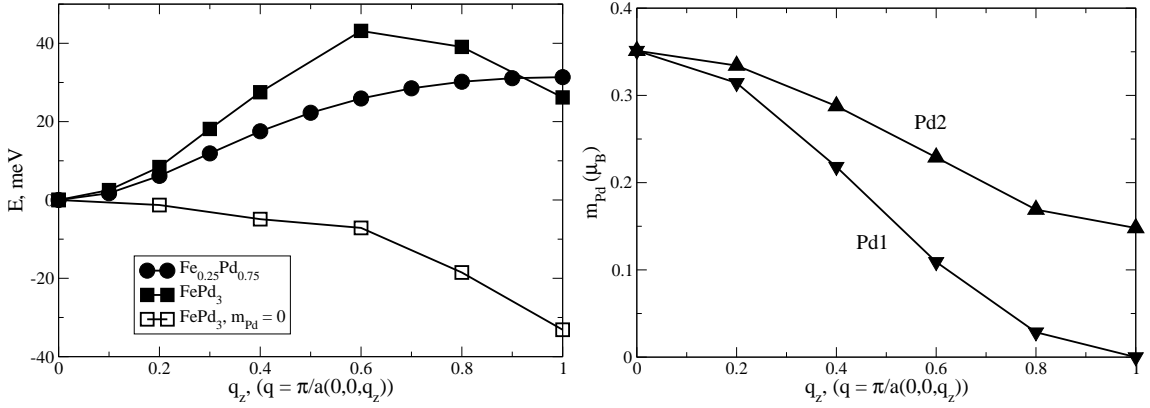

\includegraphics[width=7.5cm,angle=0]{Fe_0.25Pd_0.75_Spin_Spiral_CMP_ordered_vs_disordered.eps}  
\includegraphics[width=7.5cm,angle=0]{FePd3_Pd_mmom_vs_q_spinspiral.eps}
\caption{\label{Fig2} Results of spin-spiral calculations for ordered
  FePd$_3$ and disordered Fe$_{0.25}$Pd$_{0.75}$: (a) the energy of spin spirals as a function
 of the wave vector $q = \frac{\pi}{a}(0,0,q_z)$, obtained for the disordered
 alloy (full circles), for the ordered compound with non-zero Pd magnetic
 moments (full squares) and for the ordered compound with Pd magnetic
 moments equal to 0 (open squares); (b) - magnetic moment of
 inequivalent Pd atoms in ordered FePd$_3$ as a function of the  wave
 vector $q_z$. Fe and Pd2 occupy the sites $(0,0,0)$ and
 $(\frac{1}{2},\frac{1}{2},0)$, respectively, Pd1 atoms occupy the sites
 $(0,\frac{1}{2},\frac{1}{2})$ and $(\frac{1}{2},0,\frac{1}{2})$. }
\end{figure}


According to the experimental findings \cite{CS65}, the alloys exhibit
ferromagnetic order for Fe concentrations above 0.1\%. 
In addition, previous experimental \cite{Ger58,CWK65,CMS65,Lon68,CT71,
  Cra60, BCTS75,AWF+88} and theoretical
\cite{Kim66,Ber81,MP82,MS93,JHZQ93,OZD86} 
 investigations on the magnetic properties of  Fe$_x$Pd$_{1-x}$ alloys
 have shown a strong host polarisation by magnetic Fe impurities leading
 to a giant magnetic moment per impurity atom, up to $12.9 \mu_B$.  
A wide-spread regime of 
magnetised Pd atoms leads to ferromagnetic order in diluted  Fe$_x$Pd$_{1-x}$
alloys despite a large distance between the magnetic Fe atoms. 

The crucial role of the induced magnetic moment on Pd for the
Fe-Fe exchange interactions can be demonstrated by an analysis of the
energy of spin 
spirals as a function of a wave vector, shown in Fig. \ref{Fig2} for the
disordered alloy Fe$_{0.25}$Pd$_{0.75}$ in comparison with the results 
for the ordered compound FePd$_{3}$. 
The calculations have been performed for spin spirals 
along the $z$ direction with the Fe magnetic moments tilted by $90^o$
with respect to the $z$-axis.  
For the ordered system an increase of the wave vector of the spin
spirals is accompanied first by an increase in energy reflecting the
stability of the ferromagnetic order in the system. 
A further increase of the wave vector above $q_z > \pi/2a$,
leads to a decrease of the energy of the spin spiral (Fig. \ref{Fig2}a). 
This behavior is 
governed by a decrease of the Pd magnetic moments at these wave vectors (see
Fig. \ref{Fig2}b), that diminishes their role in the Fe-Fe exchange.

The role of Pd becomes clearly visible for the spin spirals in
FePd$_{3}$ with the  Fe-Pd exchange interactions being suppressed. This
can be achieved  
by forcing the Pd induced magnetic moments to be perpendicular to the Fe
magnetic moments, and therefore to be equal to 0 (see Appendix). In
this case the minimum of the spin-spiral energy corresponds to an AFM
state, i.e. at $q = \pi/a$, that originates from Fe-Fe exchange
interaction (see Fig. \ref{Fig1}).  
In the case of the
disordered alloy, the dependence of the spin-spiral energy on the wave vector is
different because the random distribution of the Fe atoms allows Fe atoms to be nearest neighbours with a strong FM
interaction. Due to this, the system retains the FM order 
at all values of wave vector $\vec{q}$.

\vspace{1cm}
\begin{figure}[h]
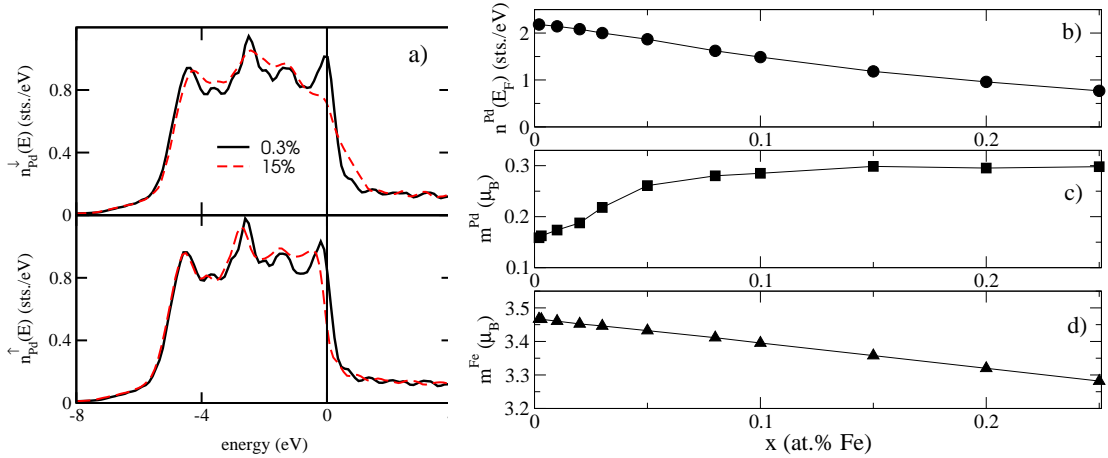

\includegraphics[width=6.cm,angle=0]{CMP_DOS.eps}
\includegraphics[width=8.5cm,angle=0]{FePd_ground_state.eps}
\caption{\label{Fig_DOS} a) Pd DOS in Fe$_x$Pd$_{1-x}$ for $x = 0.003$ and $x =
  0.15$ for spin-up (upper panel) and spin-down (lower panel) states.
Ground state characteristics of
  Fe$_x$Pd$_{1-x}$ alloys vs. Fe concentration: b) density of states of
  Pd at the Fermi level; c) Pd spin magnetic moments; d) Fe spin
  magnetic moments.  
}
\end{figure}

While the ground-state magnetic
properties of  Fe$_x$Pd$_{1-x}$ alloys can essentially be understood on the basis of
{\em ab-initio} electronic structure calculations, a theoretical
description of finite-temperature properties 
faces many difficulties due to the itinerant-electron nature
of magnetism.
Theoretical
investigations based on the results of {\em ab-initio} electronic structure
calculations have been performed for example by Mohn and Schwarz \cite{MS93}. They
used the model approach  
formulated by Bloch et al. \cite{BESV75} to describe the magnetic
behavior of a system characterised by the coexistent local- and
itinerant-electron magnetism. Within this approach the system is
characterised by two interacting subsystems: (i) one having local magnetic
moments showing a Curie-Weiss-like behaviour and (ii) an
 itinerant electron subsystem magnetically polarised by the effective
 Weiss field, with the corresponding parameters found by {\em ab-initio}
 electronic structure calculations.

The spin moment on every Pd atom is induced by the magnetic moment
of the Fe atoms and all 
surrounding induced Pd magnetic moments (see Appendix). 
The rather big absolute value and large region of Pd induced magnetic moments 
around an Fe atom is a result of
the high magnetic spin susceptibility of pure Pd and Fe$_x$Pd$_{1-x}$
alloys with small 
 Fe concentration, that is determined by a large Pd density of
 states (DOS) at the Fermi level, $n(E_F)$
  (see Fig. \ref{Fig_DOS}a). In turn, $n(E_F)$ decreases with the increase
 of Fe content in the alloy (Fig. \ref{Fig_DOS}b), resulting in
a decrease of the partial magnetic susceptibility of the Pd atoms.
Thus, at very
small Fe concentrations the induced Pd spin moment can extend to big
distances - inducing shell-by-shell a spin moment in the Pd subsystem. This polarisation mechanism decays with the distance from the magnetic impurity.
When the Fe concentration increases, the regions with the induced moments 
overlap and, as was pointed in M. Shimizu et al. \cite{SK68,TS65}, when the Fe
concentration is larger than 0.1\% the Pd magnetic properties can be
described well by band calculations, as done in our present 
calculations using the CPA alloy theory.

Figure \ref{Fig_DOS}, c and d, represent the spin magnetic moments of
Pd and Fe versus the Fe content in Fe$_x$Pd$_{1-x}$. As is seen, the Fe
magnetic moments 
change only slowly with the increase of Fe concentration, while the
variation of the Pd spin magnetic moments is rather pronounced. This can be
directly connected to the decrease of the Pd DOS at the Fermi level.

\vspace{1cm}
\begin{figure}[h]
\includegraphics[width=8.5cm,angle=0]{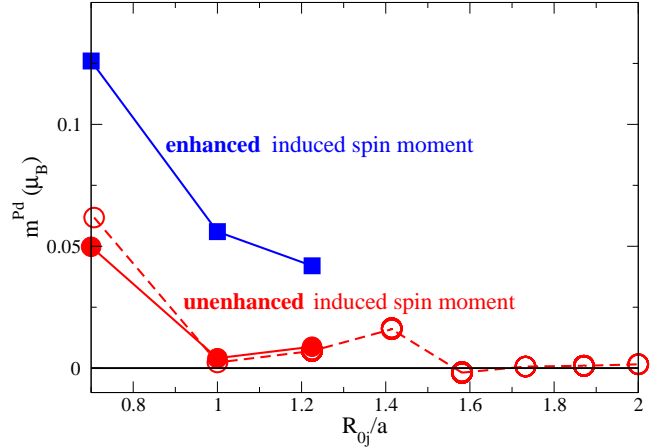}
\caption{\label{Clu_siscept} Magnetic moment distribution in Pd 
  around a single Fe impurity, as a function of the distance from the
  magnetic atom. Full squares represent the Pd magnetic moments
  self-consistently obtained within a cluster with 3 atomic shells of
  Pd around an Fe impurity embedded into a Pd host, full circles give
  the results obtained for the 
  same system but with the exchange potential switched off. The open
  circles represent the induced magnetic moments in Pd 
  calculated within linear response formalism using Eq. (\ref{Delta_m}) in
 the Appendix. 
 }
\end{figure}

For a more detailed analysis, we investigated the properties of the induced Pd
magnetic moment using the {\em ab-initio} calculations.  
In particular, we studied the distribution of the Pd magnetic moment in
a Pd host 
in the limit of very small Fe concentrations, i.e. around a single
 Fe impurity. 
To see that the induced magnetic moment at every Pd atom is determined
not only by the Fe magnetic moment but also by the surrounding
Pd magnetic moments, one can compare 
the un-enhanced induced Pd spin moments created
by only one Fe atom with the total induced spin moments in Pd.
The total induced magnetic moment distribution in Pd can be found by solving
the system of equations (\ref{Minduced2}) within a selected region around an
Fe atom (see Appendix). In addition, in   
the present work the moment distribution has been obtained
by self-consistent 
electronic structure calculations instead of using linear response formalism. 
Fig. \ref{Clu_siscept} shows the slow decay of the induced magnetic moment
with the distance (full squares). The corresponding 
un-enhanced induced moment in these calculations has been obtained
by suppressing the effective exchange B-field for the Pd atoms during the
SCF-cycle. These un-enhanced magnetic moments compare very well with
those obtained from linear response formalism (Eq. (\ref{Delta_m}))
(open circles). These are shown in Fig. \ref{Clu_siscept} also for
larger distances.  
One can see that the decrease with the distance of the 
 Pd un-enhanced spin moment is very fast compared to the enhanced one. 
For the nearest Pd neighbors of Fe atom these values differ approximately
by a factor of 2, while the difference for the next nearest neighbors is
already an order of magnitude. The local exchange enhancement, well
approximated within the linear approach at
small values of the induced magnetic moments, should keep the ratio of these
two values approximately constant. The obtained results give evidence 
for a more complicated picture of the creation of the induced magnetic
moment in accordance to the description given in the Appendix.

\begin{figure}[h]
\includegraphics[width=8.5cm,angle=0]{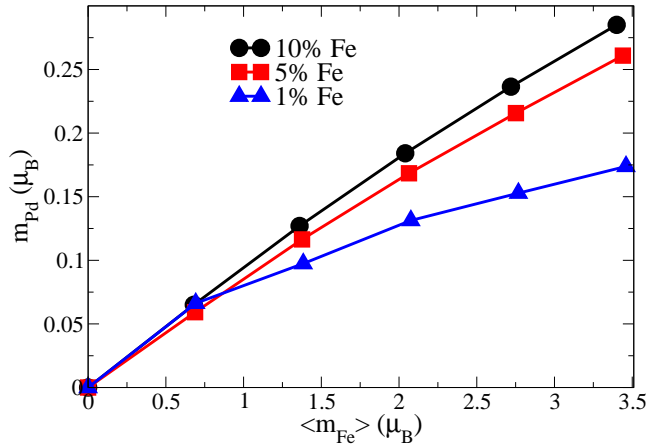}
\caption{\label{M_Pd_vs_M_Fe} Results of {\em ab-initio} calculations for
 induced Pd moment in three Fe-Pd alloys, using the
  uncompensated DLM approximation as a function
  of the Fe average magnetic moment.
}
\end{figure}

The effect of temperature-induced magnetic disorder within the
Fe-subsystem was analysed within {\em ab-initio} calculations, describing
 magnetic disorder within the uncompensated Disordered Local Moment
 (DLM) approximation. Using this approximation an effective
alloy of two types of Fe atoms with opposite spin directions and
having different concentrations is treated using the CPA alloy
theory. In this way one can study the dependence of the 
induced magnetic moment of individual Pd atoms on the average magnetic
moment in the system.
 Fig. \ref{M_Pd_vs_M_Fe} shows that the magnetic disorder in the Fe
subsystem (assumed to be temperature induced) is 
 accompanied by a decease of the total magnetic moment in the system and
 results in a decease of the induced magnetic moment in the Pd subsystem.
 One can see a rather good linear  
dependence of the induced Pd magnetic moment as a function of the
magnetic moment 
of the Fe subsystem, for nearly all Fe concentrations. Only in the limiting
case of low Fe concentration (1\%), a noteworthy deviation from linear
behavior is 
observed. This deviation will influence the final results in 
a Curie temperature evaluation correspondingly.

\vspace{1cm}

\subsection{Finite-temperature magnetism of  Fe$_x$Pd$_{1-x}$ alloys}
\label{}

\begin{figure}[h]
\includegraphics[width=3.5cm,angle=0]{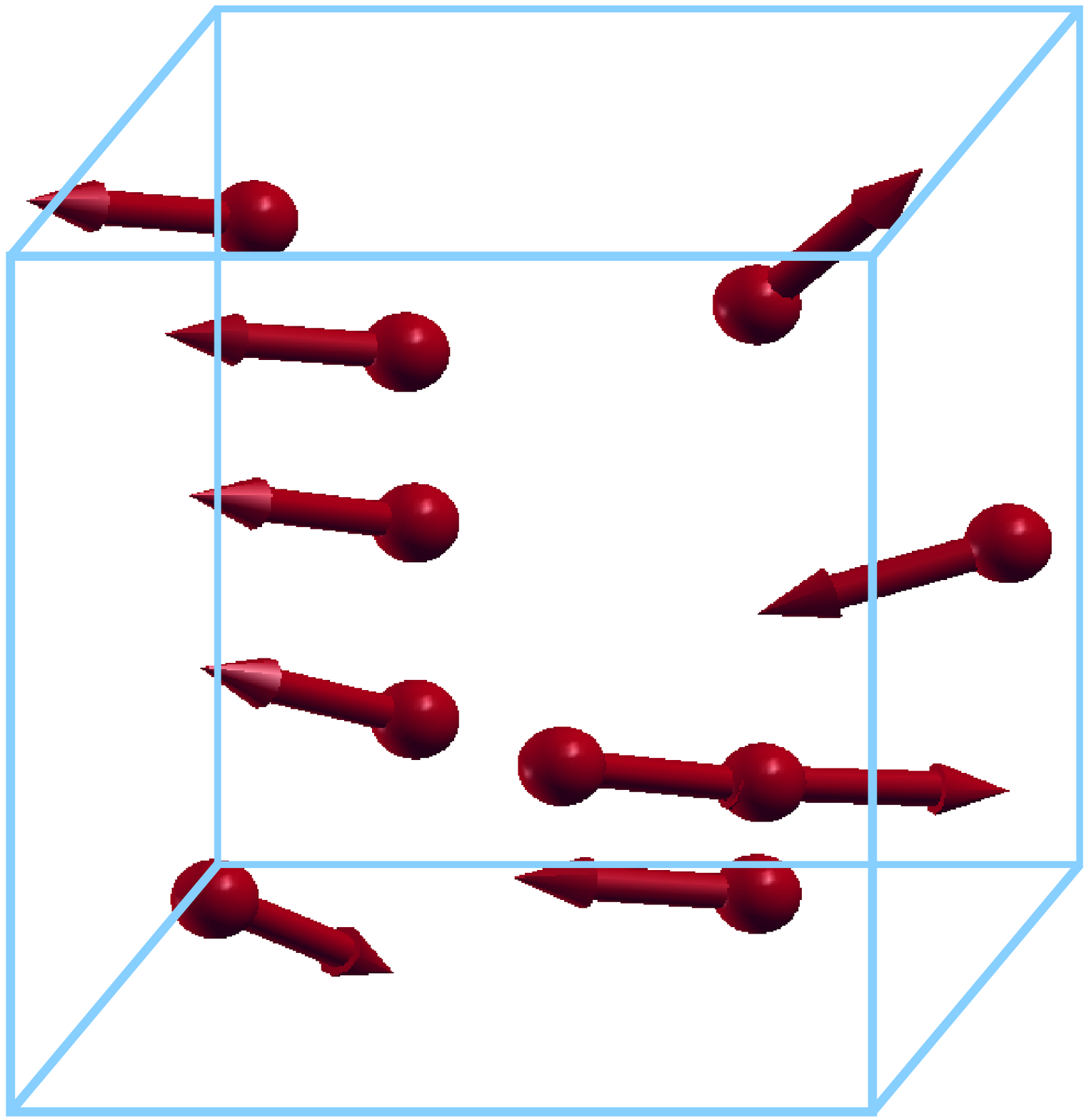}a  
\includegraphics[width=3.5cm,angle=0]{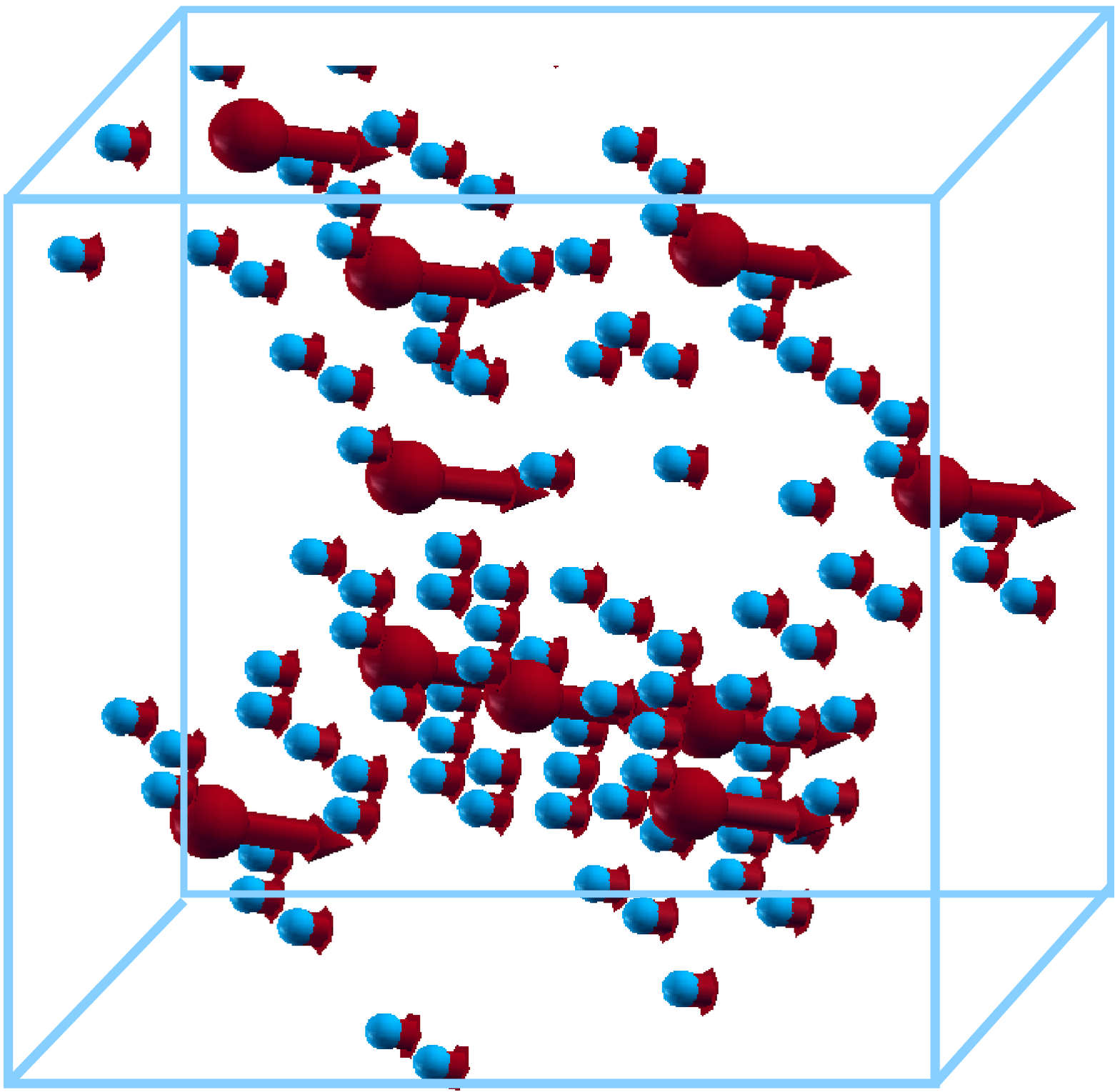}b  
\caption{\label{T_C2} The
  distribution of spin moments in unit cell used within the MC
  simulations without (a) and
  with (b) the induced Pd moments taken into account. Large arrows 
  correspond to the Fe magnetic moments, small arrows to Pd magnetic
  moments. 
}
\end{figure}

The temperature dependent magnetic properties of  Fe$_x$Pd$_{1-x}$ alloys were
investigated by performing Monte Carlo simulations. They show an absence of
an ordered FM state for the alloys with Fe concentration up to $x \approx
0.2$ if the Fe-Fe exchange interactions mediated by Fe-Pd interactions
are neglected. This clearly demonstrates the importance of these
interactions. The Curie 
temperature for disordered Fe$_{0.2}$Pd$_{0.8}$ alloy 
in this case is around 60K, much lower than observed experimentally
(around 400 K).  
To illustrate the role of Fe-Pd interactions in the formation of magnetic
order, Fig. \ref{T_C2} shows spin configurations obtained within
MC simulations for ($T = 0.1 K$) for the  Fe$_x$Pd$_{1-x}$
alloy with $x = 0.02$. Fig.  \ref{T_C2}a represents a result for the
case that the Fe-Pd exchange interactions are neglected and therefore it shows
only the magnetic moments of Fe.  Fig. \ref{T_C2}b represents the spin
configuration when the  Fe-Pd first-neighbour interactions are taken
into account and shows only those Fe and Pd magnetic moments which give a
contribution to the total energy of a system given by Eq. (\ref{Heisenberg}).

\begin{figure}[h]
\includegraphics[width=8.5cm,angle=0]{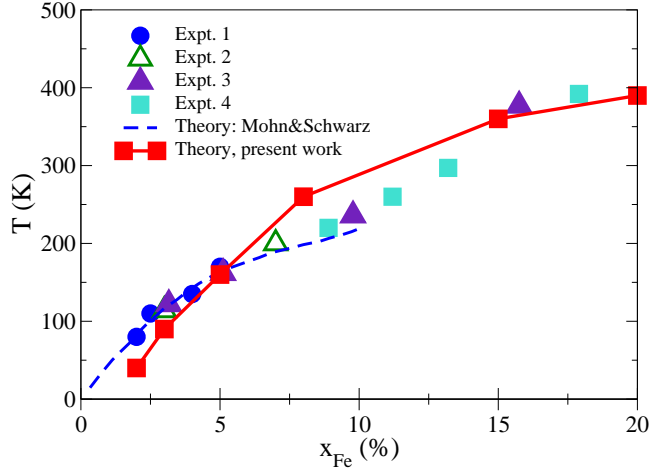}
\caption{\label{T_C} (a) The Curie temperature for different Fe
  concentrations in  Fe$_x$Pd$_{1-x}$ alloys: present results vs theoretical results of Mohn and Schwarz \cite{MS93} and  
  experimental data: Expt. 1 --
 present work, Expt. 2 \cite{CWK65}, Expt. 3 \cite{Cra60},
 Expt. 4 \cite{YCTF75}. 
}
\end{figure}

A comparison of $T_C$ obtained within the MC simulation based on the
Hamiltonian in Eq. (\ref{Heisenberg}) with the experimental data for
Fe$_x$Pd$_{1-x}$ alloys is shown in Fig. \ref{T_C}. Obviously, a
 rather good
agreement is obtained for the whole
concentration range. It should be emphasized once more that all
parameters for the model Hamiltonian (Eq. (\ref{Heisenberg})) are obtained
within {\em ab-initio} electronic structure calculations using the
scheme described in the Appendix. Of course, going beyond the various
approximations the 
final results can be improved to get better agreement with the
experimental results.  
Fig. \ref{T_C} shows that the theoretical result obtained by Mohn
and Schwarz 
\cite{MS93} for Fe$_x$Pd$_{1-x}$ alloys at small Fe concentrations is
also in a good agreement with experiment. However, it should be
emphasized that this work is based on a semi-phenomenological approach.

\vspace{1cm}

\vspace{1cm}

\section{Results for Co$_x$Pt$_{1-x}$ alloys}

\subsection{Magnetic moments and exchange coupling constants}

\begin{table}
\begin{tabular}{llccc}
\hline
\hline
system  &  & $a$\ 
        & $ \mu_{\mathrm{spin}}$\ (Co)  
        & $ \mu_{\mathrm{spin}}$\ (Pt) \\  
  &  & a.u. & $\mu_{B}$ & $\mu_{B}$  \\  
\hline
Co$_{3}$Pt & ordered     & 6.98 & 1.82 &  0.36  \\
           & disordered  & 7.06 & 1.88 &  0.25  \\ [0.5ex]
CoPt       & ordered     & 7.23 & 1.93 &  0.37   \\
           & disordered  & 7.23 & 2.03 &  0.27   \\ [0.5ex]
CoPt$_{3}$ & ordered     & 7.31 & 1.76 &  0.26   \\
           & disordered  & 7.37 & 2.19 &  0.25   \\
\hline
\hline
\end{tabular}
\caption{Equilibrium lattice constants and magnetic moments at Co and
  Pt atoms for ordered and disordered Co$_x$Pt$_{1-x}$ alloys.}
\label{tab-mom}
\end{table}

The calculated equilibrium lattice constants for ordered and disordered
Co$_x$Pt$_{1-x}$ alloys are 
shown in Tab.\ \ref{tab-mom}.  For ordered CoPt, we used a simplified
$L1_{0}$\ geometry, assuming $c$=$a$\ instead of $c$=0.98$a$\ found in
experiment.  The magnetic moments of the Co and Pt atoms for the
equilibrium lattice 
constants obtained in the scalar-relativistic mode are presented in Table\
\ref{tab-mom}.  More details about
ground-state properties of Co-Pt can be found in our earlier study
 \cite{SMME08}.

The exchange coupling constants $J_{ij}$ for the investigated systems were evaluated via
Eq.\ (\ref{eq_jij}).  The dependence of $J_{ij}$ on the
distance between the atoms $i$\ and $j$\ is displayed in
Fig.\ \ref{fig-jxc} with the left panels showing the situation when both
$i$\ and 
$j$\ are Co atoms and the right panels showing the situation when $i$\ is a Co
atom and $j$\ is a Pt atom.  If experimental lattice constants were
used instead of equilibrium lattice constants, the $J_{ij}$ constants
would change slightly but both the trends and the values would remain
similar as in Fig.\ \ref{fig-jxc}.

One can see from Fig.\ \ref{fig-jxc} that for Co$_{3}$Pt and CoPt the coupling between the moments
on the Co atoms do not differ very much from the results for their disordered counter parts, Co$_{0.75}$Pt$_{0.25}$ and Co$_{0.50}$Pt$_{0.50}$ respectively. However, the situation changes dramatically for
CoPt$_{3}$ (lower left panel in Fig.\ \ref{fig-jxc}). The pronounced difference between the data for the ordered and the disordered
system stems mainly from the fact that there are no Co atoms present for
some coordination spheres around a central Co atom in ordered
CoPt$_{3}$.  Concerning the coupling between moments on Co and Pt atoms,
the degree of long-range order has a larger influence than for the
Co--Co coupling.  For ordered alloys, the $J_{ij}$ constants significantly
vary also with composition.  For disordered alloys, on the other hand, the
$J_{ij}$ constants do not vary very much with composition.
For ordered CoPt$_{3}$, there is a surprisingly strong Co-Co coupling
between atoms which are 2.83~$a$\ apart.  We verified that for larger
distances no comparable strong coupling occurs.

\subsection{Curie temperatures}

\begin{table}
\begin{tabular}{llrr}
\hline
\hline
system  & model & $T_{C}$\ [K] & $T_{C}$\ [K] \\
        &          & ordered      & disordered   \\
\hline
fcc Co      &  MC               & 1100  &   \\
            &  experiment       & 1388  &   \\
            &  Mohn-Wohlfarth   & 3523  &   \\ [0.5ex]
Co$_{3}$Pt  &  MC, Co-Co only   & 800 & 750 \\
            &  MC, Co-Co and Co-Pt  & 900 &  880 \\
            &  experiment       & -- &  1100 \\   
            &  Mohn-Wohlfarth   & 1803  & 1120 \\ [0.5ex]
CoPt        &  MC, Co-Co only       & 360 &  620 \\
            &  MC, Co-Co and Co-Pt  & 620 &  760 \\
            &  experiment           & 727 &  830 \\
            &  Mohn-Wohlfarth   & 1964 &  850 \\ [0.5ex]
CoPt$_{3}$  &  MC, Co-Co only       & $-$180 &  370 \\
            &  MC, Co-Co and Co-Pt  & 150 &  520 \\         
            &  experiment       & 288 & 468 \\
            &  Mohn-Wohlfarth   & 241 & 510 \\
\hline
\hline
\end{tabular}
\caption{Curie temperatures $T_{C}$ for fcc Co and for ordered and
  disordered Co$_x$Pt$_{1-x}$ alloys.  Experimental results \cite{Dah85,CDK86}
  are shown together with results of our Monte-Carlo calculations with
  either both Co-Co and Co-Pt coupling or with only Co-Co coupling
  included.  For comparison, results obtained by relying on the
  Mohn-Wohlfarth theory are also shown
  \cite{QSC94,KGS99,GCS01}. Negative $T_{C}$\ implies
  antiferromagnetic ordering.}
\label{tab-tc}
\end{table}

The Curie temperatures $T_{C}$ of the investigated Co-Pt systems evaluated by means of the Monte-Carlo
technique are shown in Tab.\ \ref{tab-tc}.  In addition, results for fcc Co are given in this table. The two
theoretical Curie temperatures correspond to two different Hamiltonians used to
describe the magnetic coupling.  The first $T_{C}$ (denoted as ``Co-Co
only'' in Tab.\ \ref{tab-tc}) corresponds to the standard Heisenberg
Hamiltonian for magnetic moments only on the Co atoms.  The second
$T_{C}$ (denoted as ``Co-Co and Co-Pt'') corresponds to the extended
Heisenberg Hamiltonian (Eq. (\ref{Heisenberg})), accounting for the coupling
between moments on Co atoms $\tilde{J}^{Co-Co}_{ij}$ as well as
for the coupling between moments on Co and Pt atoms
$\tilde{J}^{Co-Pt}_{ij}$, with the moments on Pt atoms determined
via Eqs.\ (\ref{A3}) and (\ref{A4}).  The values of $T_{C}$
obtained 
earlier by relying on the semi-empirical Mohn-Wohlfarth theory
\cite{MW87} were taken from Qi et al. \cite{QSC94} for fcc Co, from
Kashyap et al. \cite{KGS99} for ordered Co-Pt compounds and from Ghosh
et al. \cite{GCS01} for disordered Co$_x$Pt$_{1-x}$ alloys.  Experimental values
\cite{Dah85,CDK86} are shown for comparison.  Note that an experimental
value for $T_{C}$ for ordered Co$_{3}$Pt is not available because the
ordered phase is not stable for this composition.

As is seen, the Mohn-Wohlfarth theory \cite{MW87} gives
reasonable agreement with experiment at small Co concentration.
However, increasing the Co content leads to large discrepancies
between the theory and experiment. This results from the
limitations of the Mohn-Wohlfarth theory: it was developed for homogeneous
itinerant-electron systems, which is not the case for Co$_x$Pt$_{1-x}$ alloys.

Our {\em ab-initio} scheme based on an extended Heisenberg Hamiltonian
(both Co-Co and Co-Pt coupling included) accounts quantitatively for
the trends of $T_{C}$ with the composition and with the degree of
long-range order.  If the coupling mediated via moments at Pt atoms is
not included, the results are unrealistic. This is especially true for
ordered CoPt$_{3}$, where an {\em antiferromagnetic} order is established
at finite temperatures (reflected by a negative value for $T_{C}$) 
 if the coupling between moments on Co and Pt is
ignored.

\section{Conclusions}

As was shown in the present work by the examples of   Fe$_x$Pd$_{1-x}$
and Co$_x$Pt$_{1-x}$ alloys, 
the finite-temperature magnetism of alloys composed of magnetic and
non-magnetic elements requires to account for the exchange interactions
between magnetic atoms, mediated by the exchange interaction with
non-magnetic atoms.
 This inplies in particular that one has to account properly for the
 induced magnetic 
moment within the Monte Carlo simulations which are based in the present
work on a corresponding extension of the standard Heisenberg Hamiltonian.
The approach presented 
suggests to describe the induced magnetic moment on non-magnetic atoms
within linear response formalism,  being proportional to the vector sum of
magnetic moments of neighboring magnetic atoms. This ansatz allows for
substantial technical simplifications and leads to
substantial improvement of the results when compared to simpler schemes.

The finite-temperature calculations for  Fe$_x$Pd$_{1-x}$ and
Co$_x$Pt$_{1-x}$ alloys performed within
this approach give the dependence of $T_{C}$ on the composition as well
as on the degree of long-range order in good agreement with
experimental data.
The case of ordered CoPt$_{3}$ also demonstrates that even if the
coupling between nearest inducing moments is antiferromagnetic, the
magnetic order can still be ferromagnetic due to the effect of
coupling between inducing and induced moments.  A mere inspection of
the $J_{ij}$ constants thus cannot serve as a reliable indicator of
ferromagnetic or antiferromagnetic order at $T \neq 0K$.

\section{Acknowledgement}

This work was supported by the DFG within the SFB 689 "Spinph\"anomene in
 reduzierten Dimensionen" as well as the
project Eb 154/20 "Spin polarisation in Heusler alloy based spintronics
systems probed by SPINAXPES", and by GA~CR within the project 202/08/0106.

\appendix
\section{Induced magnetic moments}\label{sec:App}

The magnetic moment induced on site $i$ (Pd or Pt) by the exchange field
$\vec{B}^{xc}_j$ due to magnetic moments at site $j$ can 
be calculated within the linear response formalism  
using  the expression \cite{DFVE01}:

\begin{eqnarray}
\label{Delta_m}
\hspace{-1cm}
 \vec{m}^0_i(\vec{r}) & = & - \frac{1}{\pi} Im \int^{E_F} dE  Tr  
 \vec{\sigma} \nonumber \\   
& &\times  \int_{\Omega_j} d^3r'
  G(\vec{r},\vec{r}\,',E) H_j^{xc}
  (\vec{r}\,')G(\vec{r}\,',\vec{r},E)  \nonumber \\ 
& = & \int_{\Omega_j} \chi_{ij}(\vec{r},\vec{r}\,') \vec{B}^{xc}_j(r')d^3r'
\end{eqnarray}

Here $H^{xc}_j(\vec{r}\,') = \vec{\sigma}\vec{B}^{xc}_j(\vec{r}\,') = 
\vec{\sigma}\vec{e}_M B^{xc}_j(\vec{r}\,')$ with 
$\vec{\sigma}$ - the matrix of Pauli matrices \cite{Ros61},
$\vec{e}_M$  the unit vector in the direction of 
spontaneous magnetic moment of atom $j$,  $B^{xc}_j(\vec{r})$ the local
exchange field at the site $j$. 
Note that, neglecting relativistic effects, the magnetic moment $\vec{m}^0_i$ induced by a neighboring magnetic atom  is parallel to 
the direction  $\vec{e}_M$ of the magnetic moment $\vec{M}_j$ of this
atom. 

The total 
induced magnetic moments on a Pd or Pt atom 
is represented as a response to the exchange field of all
surrounding atoms by the following expression:

\begin{eqnarray}
\label{Minduced}
\vec{m}_i(\vec{r}) & = & \sum_{j\in M} \int_{\Omega_j^M} \chi^{m-M}_{ij}(\vec{r},\vec{r}\,')
\vec{B}^{xc,M}_j(\vec{r}\,')d^3r'   \\ 
&& +\sum_{j\in m} \int_{\Omega_j^m}  \chi^{m-m}_{ij}(\vec{r},\vec{r}\,')
\vec{B}^{xc,m}_j(\vec{r}\,')d^3r'  \nonumber \\
 && + \chi^0_i(\vec{r}) \vec{B}^{xc,m}_i(\vec{r}) \;. \nonumber 
\end{eqnarray}
Analogous to $\sum_{j\in M}$\ defined at Eq.\ (\ref{A3}), the sum
$\sum_{j\in m}$\ means summation over sites with induced magnetic
moments. 

This can be reformulated in terms of local magnetic moments $M_i$ of
Fe (Co) and $m_i$ of Pd (Pt)

\begin{eqnarray}
\label{Minduced2}
\hspace{-1cm}
m_i & = & 
  \sum_{j\in M} \tilde{\chi}^{m-M}_{ij} \vec{M}_j
+ \sum_{j \in m} \tilde{\chi}^{m-m}_{ij} \vec{m}_j 
 + \tilde{\chi}_{ii} \vec{m}_i
\end{eqnarray}

with 

\begin{eqnarray}
\label{BXC_ind_suppl}
\vec{M}_i =
\int_{\Omega_{WS}} d^3r \vec{M}_i (\vec{r}); \;\;\;\; \vec{m}_i =
\int_{\Omega_{WS}} d^3r \vec{m}_i (\vec{r})\;.
\end{eqnarray}

Here we use the following reformulation for the first term in
Eq. (\ref{Minduced}), that is more convenient for the
model implementation: 

\begin{eqnarray}
\label{chi_m}
\hspace{-1cm}
&& \int_{\Omega_j^M} \chi^{m-M}_{ij}(r,r') \vec{B}^{xc,M}_j(r')d^3r' \\
&& =  \vec{M}_j \int_{\Omega_j^M} \chi^{m-M}_{ij}(r,r')
\frac{\vec{B}^{xc,M}_j(r')}{M_j}d^3r' \nonumber \\
&& = \tilde{\chi}^{m-M}_{ij}
\vec{M}_j   \nonumber
\end{eqnarray}

For the second term, using a linearised expression for the exchange
potential in the case of a small induced magnetic moment on Pd and Pt
sites \cite{DFVE01, ME06}, one can write analogously:

\begin{eqnarray}
\label{chi_mind}
\hspace{-1cm}
&& \int_{\Omega_j^m} \chi^{m-m}_{ij}(r,r') \vec{B}^{xc,m}_j(r')d^3r'
 \\
 & = & \vec{m}_j \int_{\Omega_j^m} \chi^{m-m}_{ij}(r,r') \frac{\delta
  V_j^{xc}[n,m]}{\delta m(\vec{r})}|_{m_{spin} = 0}
\frac{m_j(\vec{r})}{m_j}d^3r'  \nonumber \\ 
&=& \tilde{\chi}^{m-m}_{ij} \vec{m}_j\,.  \nonumber
\end{eqnarray}

Solving the system of equations (\ref{Minduced2})  for a restricted
region around a magnetic impurity atom gives the distribution of the
induced magnetic moments on the
non-magnetic atoms. This can be done for the ground state ($T = 0$K). 
Alternatively, without any approximations, one can get these values within
{\em ab-initio} calculations for embedded magnetic atoms by using the CPA alloy
theory, assuming an uniform distribution of the induced magnetic moment.

Strictly spoken, one can go beyond the  linear approximation in the expansion
of the exchange potential. However, the linear approximation 
makes the use of this scheme in subsequent Monte Carlo simulations much
easier.

By making an additional simplification one can restrict to one response
function $X^{m-M}_{ij}$ within the Monte-Carlo simulations.
This quantity is defined to give the induced magnetic moment as a
response  to the exchange fields of only the surrounding nearest
neighbour magnetic atoms:

\begin{eqnarray}
\label{Minduced3}
\hspace{-1cm}
\vec{m}_i & = & 
  \sum_{j \in M} X^{m-M}_{ij} \vec{M}_j.
\end{eqnarray}

\vspace{1cm}
\begin{figure}
\includegraphics[viewport=0.8cm 0.8cm 16.0cm 20.0cm]{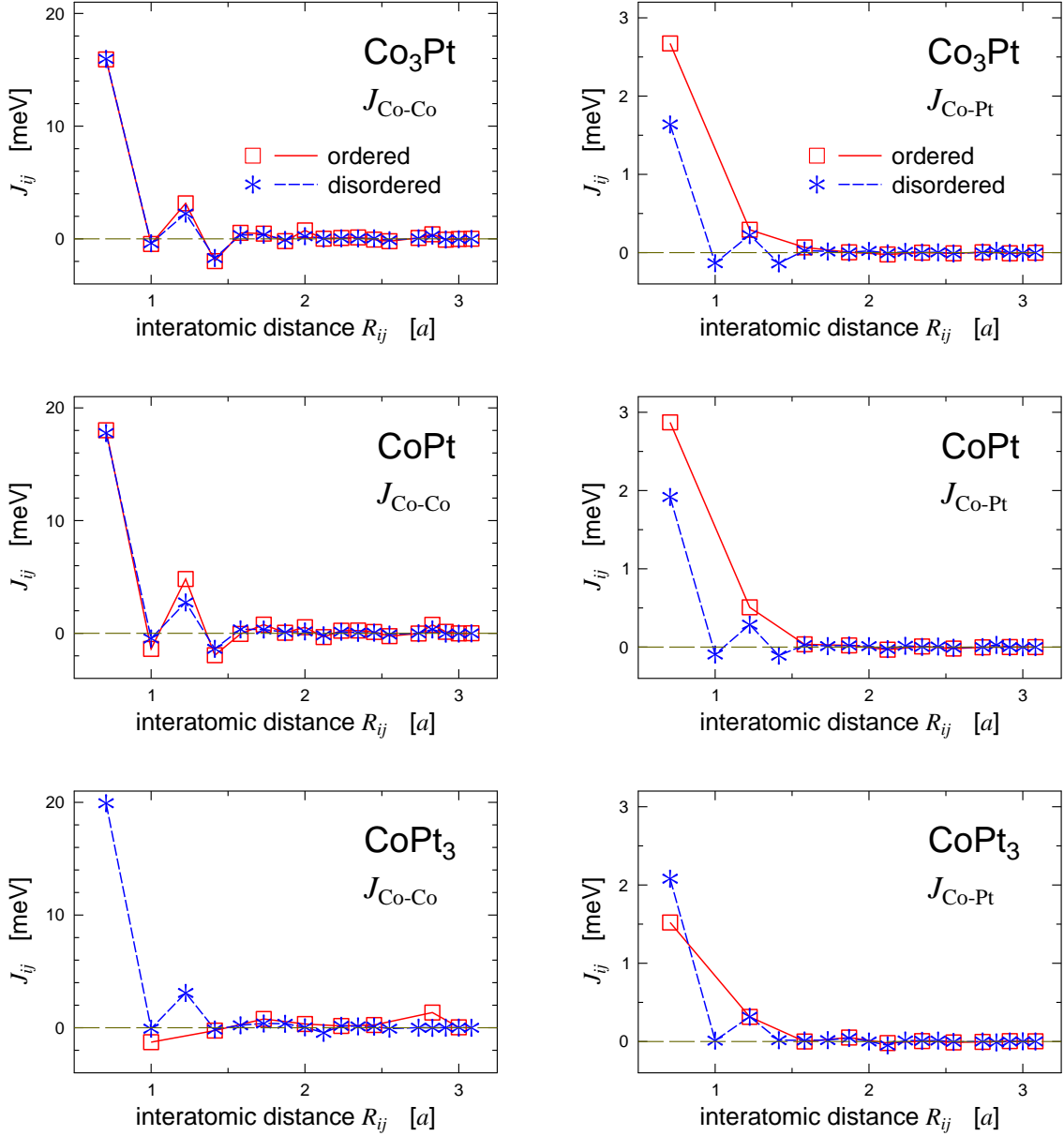}%
\caption{Exchange coupling constants $J_{ij}$ between moments of a Co atom on site $i$ and on other
  surrounding Co atoms ($j$) (left panels) and between moments on a Co atom ($i$) and on surrounding Pt atoms ($j$) (right panel), for ordered and disordered
  Co$_x$Pt$_{1-x}$ alloys. The horizontal axis shows the distance
  $R_{ij}$ between $i$\ and $j$\ atoms in units of lattice
  constants.}
\label{fig-jxc}
\end{figure}


\begin{thebibliography}{54}
\expandafter\ifx\csname natexlab\endcsname\relax\def\natexlab#1{#1}\fi
\expandafter\ifx\csname bibnamefont\endcsname\relax
  \def\bibnamefont#1{#1}\fi
\expandafter\ifx\csname bibfnamefont\endcsname\relax
  \def\bibfnamefont#1{#1}\fi
\expandafter\ifx\csname citenamefont\endcsname\relax
  \def\citenamefont#1{#1}\fi
\expandafter\ifx\csname url\endcsname\relax
  \def\url#1{\texttt{#1}}\fi
\expandafter\ifx\csname urlprefix\endcsname\relax\def\urlprefix{URL }\fi
\providecommand{\bibinfo}[2]{#2}
\providecommand{\eprint}[2][]{\url{#2}}

\bibitem[{\citenamefont{K{\"u}bler}(2000)}]{Kueb00}
\bibinfo{author}{\bibfnamefont{J.}~\bibnamefont{K{\"u}bler}},
  \emph{\bibinfo{title}{Theory of itinerant electron magnetism}}
  (\bibinfo{publisher}{Oxford University Press}, \bibinfo{address}{Oxford},
  \bibinfo{year}{2000}), p. \bibinfo{pages}{460}.

\bibitem[{\citenamefont{Mohn}(2003)}]{Mohn03}
\bibinfo{author}{\bibfnamefont{P.}~\bibnamefont{Mohn}},
  \emph{\bibinfo{title}{Magnetism in the Solid State}}
  (\bibinfo{publisher}{Springer}, \bibinfo{address}{Berlin},
  \bibinfo{year}{2003}), p. \bibinfo{pages}{215}.

\bibitem[{\citenamefont{Liechtenstein et~al.}(1987)\citenamefont{Liechtenstein,
  Katsnelson, Antropov, and Gubanov}}]{LKAG87}
\bibinfo{author}{\bibfnamefont{A.~I.} \bibnamefont{Liechtenstein}},
  \bibinfo{author}{\bibfnamefont{M.~I.} \bibnamefont{Katsnelson}},
  \bibinfo{author}{\bibfnamefont{V.~P.} \bibnamefont{Antropov}},
  \bibnamefont{and} \bibinfo{author}{\bibfnamefont{V.~A.}
  \bibnamefont{Gubanov}}, \bibinfo{journal}{J. Magn. Magn. Materials}
  \textbf{\bibinfo{volume}{67}}, \bibinfo{pages}{65} (\bibinfo{year}{1987}).

\bibitem[{\citenamefont{Fischer et~al.}(2009)\citenamefont{Fischer, D\"ane,
  Ernst, Bruno, L\"uders, Szotek, Temmerman, and Hergert}}]{FDE09}
\bibinfo{author}{\bibfnamefont{G.}~\bibnamefont{Fischer}},
  \bibinfo{author}{\bibfnamefont{M.}~\bibnamefont{D\"ane}},
  \bibinfo{author}{\bibfnamefont{A.}~\bibnamefont{Ernst}},
  \bibinfo{author}{\bibfnamefont{P.}~\bibnamefont{Bruno}},
  \bibinfo{author}{\bibfnamefont{M.}~\bibnamefont{L\"uders}},
  \bibinfo{author}{\bibfnamefont{Z.}~\bibnamefont{Szotek}},
  \bibinfo{author}{\bibfnamefont{W.}~\bibnamefont{Temmerman}},
  \bibnamefont{and} \bibinfo{author}{\bibfnamefont{W.}~\bibnamefont{Hergert}},
  \bibinfo{journal}{Phys. Rev. B} \textbf{\bibinfo{volume}{80}},
  \bibinfo{pages}{014408} (\bibinfo{year}{2009}).

\bibitem[{\citenamefont{Fukushima et~al.}(2004)\citenamefont{Fukushima, Sato,
  Katayama-Yoshida, and Dederichs}}]{FSK04}
\bibinfo{author}{\bibfnamefont{T.}~\bibnamefont{Fukushima}},
  \bibinfo{author}{\bibfnamefont{K.}~\bibnamefont{Sato}},
  \bibinfo{author}{\bibfnamefont{H.}~\bibnamefont{Katayama-Yoshida}},
  \bibnamefont{and} \bibinfo{author}{\bibfnamefont{P.~H.}
  \bibnamefont{Dederichs}}, \bibinfo{journal}{Jap. J. Appl. Phys.}
  \textbf{\bibinfo{volume}{43}}, \bibinfo{pages}{L1416} (\bibinfo{year}{2004}).

\bibitem[{\citenamefont{Turek et~al.}(2003)\citenamefont{Turek, Kudrnovsk\'y,
  Drchal, Bruno, and Bl\"ugel}}]{TKDB03}
\bibinfo{author}{\bibfnamefont{I.}~\bibnamefont{Turek}},
  \bibinfo{author}{\bibfnamefont{J.}~\bibnamefont{Kudrnovsk\'y}},
  \bibinfo{author}{\bibfnamefont{V.}~\bibnamefont{Drchal}},
  \bibinfo{author}{\bibfnamefont{P.}~\bibnamefont{Bruno}}, \bibnamefont{and}
  \bibinfo{author}{\bibfnamefont{S.}~\bibnamefont{Bl\"ugel}},
  \bibinfo{journal}{phys. stat. sol. (b)} \textbf{\bibinfo{volume}{236}},
  \bibinfo{pages}{318} (\bibinfo{year}{2003}).

\bibitem[{\citenamefont{Hubbard}(1979)}]{Hub79}
\bibinfo{author}{\bibfnamefont{J.}~\bibnamefont{Hubbard}},
  \bibinfo{journal}{Phys. Rev. B} \textbf{\bibinfo{volume}{19}},
  \bibinfo{pages}{2626} (\bibinfo{year}{1979}).

\bibitem[{\citenamefont{Moriya}(1985)}]{Mor85}
\bibinfo{author}{\bibfnamefont{T.}~\bibnamefont{Moriya}},
  \emph{\bibinfo{title}{Spin Fluctuations in Itinerant Electron Magnetism}}
  (\bibinfo{publisher}{Springer}, \bibinfo{address}{Berlin},
  \bibinfo{year}{1985}).

\bibitem[{\citenamefont{Hasegava}(1979)}]{Has79}
\bibinfo{author}{\bibfnamefont{H.}~\bibnamefont{Hasegava}},
  \bibinfo{journal}{J. Phys. Soc. Japan} \textbf{\bibinfo{volume}{46}},
  \bibinfo{pages}{1504} (\bibinfo{year}{1979}).

\bibitem[{\citenamefont{Korenman et~al.}(1977)\citenamefont{Korenman, Murray,
  and Prange}}]{KMP77}
\bibinfo{author}{\bibfnamefont{V.}~\bibnamefont{Korenman}},
  \bibinfo{author}{\bibfnamefont{J.~L.} \bibnamefont{Murray}},
  \bibnamefont{and} \bibinfo{author}{\bibfnamefont{R.~E.}
  \bibnamefont{Prange}}, \bibinfo{journal}{Phys. Rev. B}
  \textbf{\bibinfo{volume}{16}}, \bibinfo{pages}{4032} (\bibinfo{year}{1977}).

\bibitem[{\citenamefont{Uhl and K\"ubler}(1996)}]{UK96}
\bibinfo{author}{\bibfnamefont{M.}~\bibnamefont{Uhl}} \bibnamefont{and}
  \bibinfo{author}{\bibfnamefont{J.}~\bibnamefont{K\"ubler}},
  \bibinfo{journal}{Phys. Rev. Lett.} \textbf{\bibinfo{volume}{77}},
  \bibinfo{pages}{334} (\bibinfo{year}{1996}),
  \urlprefix\url{http://link.aps.org/doi/10.1103/PhysRevLett.77.334}.

\bibitem[{\citenamefont{Rosengaard and Johansson}(1997)}]{RJ97}
\bibinfo{author}{\bibfnamefont{N.~M.} \bibnamefont{Rosengaard}}
  \bibnamefont{and}
  \bibinfo{author}{\bibfnamefont{B.}~\bibnamefont{Johansson}},
  \bibinfo{journal}{Phys. Rev. B} \textbf{\bibinfo{volume}{55}},
  \bibinfo{pages}{14975} (\bibinfo{year}{1997}).

\bibitem[{\citenamefont{Ruban et~al.}(2007)\citenamefont{Ruban, Khmelevskyi,
  Mohn, and Johansson}}]{RKMJ07}
\bibinfo{author}{\bibfnamefont{A.~V.} \bibnamefont{Ruban}},
  \bibinfo{author}{\bibfnamefont{S.}~\bibnamefont{Khmelevskyi}},
  \bibinfo{author}{\bibfnamefont{P.}~\bibnamefont{Mohn}}, \bibnamefont{and}
  \bibinfo{author}{\bibfnamefont{B.}~\bibnamefont{Johansson}},
  \bibinfo{journal}{Phys. Rev. B} \textbf{\bibinfo{volume}{75}},
  \bibinfo{pages}{054402} (\bibinfo{year}{2007}).

\bibitem[{\citenamefont{Williams et~al.}(1981)\citenamefont{Williams, Zeller,
  Moruzzi, Gelatt, and Kubler}}]{WZM+81}
\bibinfo{author}{\bibfnamefont{A.~R.} \bibnamefont{Williams}},
  \bibinfo{author}{\bibfnamefont{R.}~\bibnamefont{Zeller}},
  \bibinfo{author}{\bibfnamefont{V.~L.} \bibnamefont{Moruzzi}},
  \bibinfo{author}{\bibfnamefont{C.~D.} \bibnamefont{Gelatt}},
  \bibnamefont{and} \bibinfo{author}{\bibfnamefont{J.}~\bibnamefont{Kubler}},
  \bibinfo{journal}{J. Appl. Phys.} \textbf{\bibinfo{volume}{52}},
  \bibinfo{pages}{2067} (\bibinfo{year}{1981}).

\bibitem[{\citenamefont{Mohn and Schwarz}(1993)}]{MS93}
\bibinfo{author}{\bibfnamefont{P.}~\bibnamefont{Mohn}} \bibnamefont{and}
  \bibinfo{author}{\bibfnamefont{K.}~\bibnamefont{Schwarz}},
  \bibinfo{journal}{J. Phys.: Condens. Matter} \textbf{\bibinfo{volume}{5}},
  \bibinfo{pages}{5099} (\bibinfo{year}{1993}).

\bibitem[{\citenamefont{Mryasov et~al.}(2005)\citenamefont{Mryasov, Nowak,
  Guslienko, and Chantrell}}]{MNG05}
\bibinfo{author}{\bibfnamefont{O.~N.} \bibnamefont{Mryasov}},
  \bibinfo{author}{\bibfnamefont{U.}~\bibnamefont{Nowak}},
  \bibinfo{author}{\bibfnamefont{K.~Y.} \bibnamefont{Guslienko}},
  \bibnamefont{and} \bibinfo{author}{\bibfnamefont{R.~W.}
  \bibnamefont{Chantrell}}, \bibinfo{journal}{Europhys. Lett.}
  \textbf{\bibinfo{volume}{69}}, \bibinfo{pages}{805} (\bibinfo{year}{2005}).

\bibitem[{\citenamefont{Mryasov}(2005)}]{Mry05}
\bibinfo{author}{\bibfnamefont{O.~N.} \bibnamefont{Mryasov}},
  \bibinfo{journal}{Phase Transitions} \textbf{\bibinfo{volume}{78}},
  \bibinfo{pages}{197} (\bibinfo{year}{2005}).

\bibitem[{\citenamefont{Lezaic et~al.}(2006)\citenamefont{Lezaic, Mavropoulos,
  Enkovaara, Bihlmayer, and Bl\"{u}gel}}]{LME+06}
\bibinfo{author}{\bibfnamefont{M.}~\bibnamefont{Lezaic}},
  \bibinfo{author}{\bibfnamefont{P.}~\bibnamefont{Mavropoulos}},
  \bibinfo{author}{\bibfnamefont{J.}~\bibnamefont{Enkovaara}},
  \bibinfo{author}{\bibfnamefont{G.}~\bibnamefont{Bihlmayer}},
  \bibnamefont{and}
  \bibinfo{author}{\bibfnamefont{S.}~\bibnamefont{Bl\"{u}gel}},
  \bibinfo{journal}{Phys. Rev. Lett.} \textbf{\bibinfo{volume}{97}},
  \bibinfo{pages}{026404} (\bibinfo{year}{2006}).

\bibitem[{\citenamefont{Sandratskii et~al.}(2007)\citenamefont{Sandratskii,
  Singer, and \c{S}a\c{s}io\u{g}lu}}]{SSS07}
\bibinfo{author}{\bibfnamefont{L.~M.} \bibnamefont{Sandratskii}},
  \bibinfo{author}{\bibfnamefont{R.}~\bibnamefont{Singer}}, \bibnamefont{and}
  \bibinfo{author}{\bibfnamefont{E.}~\bibnamefont{\c{S}a\c{s}io\u{g}lu}},
  \bibinfo{journal}{Phys. Rev. B} \textbf{\bibinfo{volume}{76}},
  \bibinfo{pages}{184406} (\bibinfo{year}{2007}).

\bibitem[{\citenamefont{Christodoulides
  et~al.}(2000)\citenamefont{Christodoulides, Huang, Zhang, Hadjipanayis,
  Panagiotopoulos, and Niarchos}}]{CHZ+00}
\bibinfo{author}{\bibfnamefont{J.~A.} \bibnamefont{Christodoulides}},
  \bibinfo{author}{\bibfnamefont{Y.}~\bibnamefont{Huang}},
  \bibinfo{author}{\bibfnamefont{Y.}~\bibnamefont{Zhang}},
  \bibinfo{author}{\bibfnamefont{G.~C.} \bibnamefont{Hadjipanayis}},
  \bibinfo{author}{\bibfnamefont{I.}~\bibnamefont{Panagiotopoulos}},
  \bibnamefont{and} \bibinfo{author}{\bibfnamefont{D.}~\bibnamefont{Niarchos}},
  \bibinfo{journal}{J. Appl. Phys.} \textbf{\bibinfo{volume}{87}},
  \bibinfo{pages}{6938} (\bibinfo{year}{2000}).

\bibitem[{\citenamefont{Moulas et~al.}(2008)\citenamefont{Moulas, Lehnert,
  Rusponi, Zabloudil, Etz, Ouazi, Etzkorn, Bencok, Gambardella, Weinberger
  et~al.}}]{MLR+08}
\bibinfo{author}{\bibfnamefont{G.}~\bibnamefont{Moulas}},
  \bibinfo{author}{\bibfnamefont{A.}~\bibnamefont{Lehnert}},
  \bibinfo{author}{\bibfnamefont{S.}~\bibnamefont{Rusponi}},
  \bibinfo{author}{\bibfnamefont{J.}~\bibnamefont{Zabloudil}},
  \bibinfo{author}{\bibfnamefont{C.}~\bibnamefont{Etz}},
  \bibinfo{author}{\bibnamefont{Ouazi}},
  \bibinfo{author}{\bibfnamefont{M.}~\bibnamefont{Etzkorn}},
  \bibinfo{author}{\bibfnamefont{P.}~\bibnamefont{Bencok}},
  \bibinfo{author}{\bibfnamefont{P.}~\bibnamefont{Gambardella}},
  \bibinfo{author}{\bibfnamefont{P.}~\bibnamefont{Weinberger}},
  \bibnamefont{et~al.}, \bibinfo{journal}{Phys. Rev. B}
  \textbf{\bibinfo{volume}{78}}, \bibinfo{pages}{214424}
  (\bibinfo{year}{2008}).

\bibitem[{\citenamefont{Kashyap et~al.}(1999)\citenamefont{Kashyap, Garg,
  Solanki, Nautiyal, and Auluck}}]{KGS99}
\bibinfo{author}{\bibfnamefont{A.}~\bibnamefont{Kashyap}},
  \bibinfo{author}{\bibfnamefont{K.~B.} \bibnamefont{Garg}},
  \bibinfo{author}{\bibfnamefont{A.~K.} \bibnamefont{Solanki}},
  \bibinfo{author}{\bibfnamefont{T.}~\bibnamefont{Nautiyal}}, \bibnamefont{and}
  \bibinfo{author}{\bibfnamefont{S.}~\bibnamefont{Auluck}},
  \bibinfo{journal}{Phys. Rev. B} \textbf{\bibinfo{volume}{60}},
  \bibinfo{pages}{2262} (\bibinfo{year}{1999}).

\bibitem[{\citenamefont{Ebert}(2000)}]{Ebe00}
\bibinfo{author}{\bibfnamefont{H.}~\bibnamefont{Ebert}}, in
  \emph{\bibinfo{booktitle}{Electronic Structure and Physical Properties of
  Solids}}, edited by
  \bibinfo{editor}{\bibfnamefont{H.}~\bibnamefont{Dreyss\'{e}}}
  (\bibinfo{publisher}{Springer}, \bibinfo{address}{Berlin},
  \bibinfo{year}{2000}), vol. \bibinfo{volume}{535} of
  \emph{\bibinfo{series}{Lecture Notes in Physics}}, p. \bibinfo{pages}{191}.

\bibitem[{\citenamefont{Vosko et~al.}(1980)\citenamefont{Vosko, Wilk, and
  Nusair}}]{VWN80}
\bibinfo{author}{\bibfnamefont{S.~H.} \bibnamefont{Vosko}},
  \bibinfo{author}{\bibfnamefont{L.}~\bibnamefont{Wilk}}, \bibnamefont{and}
  \bibinfo{author}{\bibfnamefont{M.}~\bibnamefont{Nusair}},
  \bibinfo{journal}{Can. J. Phys.} \textbf{\bibinfo{volume}{58}},
  \bibinfo{pages}{1200} (\bibinfo{year}{1980}).

\bibitem[{\citenamefont{Weinberger}(1990)}]{Wei90}
\bibinfo{author}{\bibfnamefont{P.}~\bibnamefont{Weinberger}},
  \emph{\bibinfo{title}{Electron Scattering Theory for Ordered and Disordered
  Matter}} (\bibinfo{publisher}{Oxford University Press},
  \bibinfo{address}{Oxford}, \bibinfo{year}{1990}).

\bibitem[{\citenamefont{Binder}(1997)}]{Bin97}
\bibinfo{author}{\bibfnamefont{K.}~\bibnamefont{Binder}},
  \bibinfo{journal}{Rep. Prog. Phys.} \textbf{\bibinfo{volume}{60}},
  \bibinfo{pages}{487} (\bibinfo{year}{1997}).

\bibitem[{\citenamefont{Landau and Binder}(2000)}]{LB00}
\bibinfo{author}{\bibfnamefont{D.~P.} \bibnamefont{Landau}} \bibnamefont{and}
  \bibinfo{author}{\bibfnamefont{K.}~\bibnamefont{Binder}},
  \emph{\bibinfo{title}{A Guide to Monte Carlo simulations in statistical
  physics}} (\bibinfo{publisher}{Cambridge University Press},
  \bibinfo{address}{Cambridge}, \bibinfo{year}{2000}).

\bibitem[{\citenamefont{Crangle and Scott}(1965)}]{CS65}
\bibinfo{author}{\bibfnamefont{J.}~\bibnamefont{Crangle}} \bibnamefont{and}
  \bibinfo{author}{\bibfnamefont{W.~R.} \bibnamefont{Scott}},
  \bibinfo{journal}{J. Appl. Phys.} \textbf{\bibinfo{volume}{36}},
  \bibinfo{pages}{921} (\bibinfo{year}{1965}).

\bibitem[{\citenamefont{Gerstenberg}(1958)}]{Ger58}
\bibinfo{author}{\bibfnamefont{V.~D.} \bibnamefont{Gerstenberg}},
  \bibinfo{journal}{Annalen der Physik} \textbf{\bibinfo{volume}{7}},
  \bibinfo{pages}{236} (\bibinfo{year}{1958}).

\bibitem[{\citenamefont{Cable et~al.}(1965)\citenamefont{Cable, Wollan, and
  Koehler}}]{CWK65}
\bibinfo{author}{\bibfnamefont{J.~W.} \bibnamefont{Cable}},
  \bibinfo{author}{\bibfnamefont{E.~O.} \bibnamefont{Wollan}},
  \bibnamefont{and} \bibinfo{author}{\bibfnamefont{W.~C.}
  \bibnamefont{Koehler}}, \bibinfo{journal}{Phys. Rev.}
  \textbf{\bibinfo{volume}{138}}, \bibinfo{pages}{A755} (\bibinfo{year}{1965}).

\bibitem[{\citenamefont{Craig et~al.}(1965)\citenamefont{Craig, Mozer, and
  Romeo}}]{CMS65}
\bibinfo{author}{\bibfnamefont{P.~P.} \bibnamefont{Craig}},
  \bibinfo{author}{\bibfnamefont{B.}~\bibnamefont{Mozer}}, \bibnamefont{and}
  \bibinfo{author}{\bibfnamefont{S.}~\bibnamefont{Romeo}},
  \bibinfo{journal}{Phys. Rev. Lett.} \textbf{\bibinfo{volume}{14}},
  \bibinfo{pages}{895} (\bibinfo{year}{1965}).

\bibitem[{\citenamefont{Longworth}(1968)}]{Lon68}
\bibinfo{author}{\bibfnamefont{G.}~\bibnamefont{Longworth}},
  \bibinfo{journal}{Phys. Rev.} \textbf{\bibinfo{volume}{172}},
  \bibinfo{pages}{572} (\bibinfo{year}{1968}).

\bibitem[{\citenamefont{Chouteau and Tournier}(1971)}]{CT71}
\bibinfo{author}{\bibfnamefont{G.}~\bibnamefont{Chouteau}} \bibnamefont{and}
  \bibinfo{author}{\bibfnamefont{R.}~\bibnamefont{Tournier}},
  \bibinfo{journal}{Journal de Physique C} \textbf{\bibinfo{volume}{32}},
  \bibinfo{pages}{1002} (\bibinfo{year}{1971}).

\bibitem[{\citenamefont{Crandle}(1960)}]{Cra60}
\bibinfo{author}{\bibfnamefont{J.}~\bibnamefont{Crandle}},
  \bibinfo{journal}{Philosophical Magazine} \textbf{\bibinfo{volume}{5}},
  \bibinfo{pages}{335} (\bibinfo{year}{1960}).

\bibitem[{\citenamefont{Yen and Chen}(1975)}]{BCTS75}
\bibinfo{author}{\bibfnamefont{B.-H.} \bibnamefont{Yen}} \bibnamefont{and}
  \bibinfo{author}{\bibfnamefont{J.}~\bibnamefont{Chen}},
  \bibinfo{journal}{Chinese Journal of Physics} \textbf{\bibinfo{volume}{13}},
  \bibinfo{pages}{1} (\bibinfo{year}{1975}).

\bibitem[{\citenamefont{van Acker et~al.}(1988)\citenamefont{van Acker, Weijs,
  Fuggle, Horn, Wilke, Haak, Saalfeld, Kuhlenbeck, Braun, Williams
  et~al.}}]{AWF+88}
\bibinfo{author}{\bibfnamefont{J.~F.} \bibnamefont{van Acker}},
  \bibinfo{author}{\bibfnamefont{P.~W.~J.} \bibnamefont{Weijs}},
  \bibinfo{author}{\bibfnamefont{J.~C.} \bibnamefont{Fuggle}},
  \bibinfo{author}{\bibfnamefont{K.}~\bibnamefont{Horn}},
  \bibinfo{author}{\bibfnamefont{W.}~\bibnamefont{Wilke}},
  \bibinfo{author}{\bibfnamefont{H.}~\bibnamefont{Haak}},
  \bibinfo{author}{\bibfnamefont{H.}~\bibnamefont{Saalfeld}},
  \bibinfo{author}{\bibfnamefont{H.}~\bibnamefont{Kuhlenbeck}},
  \bibinfo{author}{\bibfnamefont{W.}~\bibnamefont{Braun}},
  \bibinfo{author}{\bibfnamefont{G.~P.} \bibnamefont{Williams}},
  \bibnamefont{et~al.}, \bibinfo{journal}{Phys. Rev. B}
  \textbf{\bibinfo{volume}{38}}, \bibinfo{pages}{10463} (\bibinfo{year}{1988}).

\bibitem[{\citenamefont{Kim}(1966)}]{Kim66}
\bibinfo{author}{\bibfnamefont{D.-K.} \bibnamefont{Kim}},
  \bibinfo{journal}{Phys. Rev.} \textbf{\bibinfo{volume}{149}},
  \bibinfo{pages}{434} (\bibinfo{year}{1966}).

\bibitem[{\citenamefont{Bergmann}(1981)}]{Ber81}
\bibinfo{author}{\bibfnamefont{G.}~\bibnamefont{Bergmann}},
  \bibinfo{journal}{Phys. Rev. B} \textbf{\bibinfo{volume}{23}},
  \bibinfo{pages}{3805} (\bibinfo{year}{1981}).

\bibitem[{\citenamefont{Medina and Parra}(1982)}]{MP82}
\bibinfo{author}{\bibfnamefont{R.}~\bibnamefont{Medina}} \bibnamefont{and}
  \bibinfo{author}{\bibfnamefont{R.~E.} \bibnamefont{Parra}},
  \bibinfo{journal}{J. Appl. Phys.} \textbf{\bibinfo{volume}{53}},
  \bibinfo{pages}{2201} (\bibinfo{year}{1982}).

\bibitem[{\citenamefont{Jian-Wang et~al.}(1993)\citenamefont{Jian-Wang, He-Lie,
  Zhi, and Qing-Qi}}]{JHZQ93}
\bibinfo{author}{\bibfnamefont{C.}~\bibnamefont{Jian-Wang}},
  \bibinfo{author}{\bibfnamefont{L.}~\bibnamefont{He-Lie}},
  \bibinfo{author}{\bibfnamefont{Z.}~\bibnamefont{Zhi}}, \bibnamefont{and}
  \bibinfo{author}{\bibfnamefont{Z.}~\bibnamefont{Qing-Qi}},
  \bibinfo{journal}{Phys. Rev. B} \textbf{\bibinfo{volume}{5}},
  \bibinfo{pages}{5343} (\bibinfo{year}{1993}).

\bibitem[{\citenamefont{Oswald et~al.}(1986)\citenamefont{Oswald, Zeller, and
  Dederichs}}]{OZD86}
\bibinfo{author}{\bibfnamefont{A.}~\bibnamefont{Oswald}},
  \bibinfo{author}{\bibfnamefont{R.}~\bibnamefont{Zeller}}, \bibnamefont{and}
  \bibinfo{author}{\bibfnamefont{P.~H.} \bibnamefont{Dederichs}},
  \bibinfo{journal}{Phys. Rev. Lett.} \textbf{\bibinfo{volume}{56}},
  \bibinfo{pages}{1419} (\bibinfo{year}{1986}).

\bibitem[{\citenamefont{Bloch et~al.}(1975)\citenamefont{Bloch, Edwards,
  Shimizu, and Voiron}}]{BESV75}
\bibinfo{author}{\bibfnamefont{D.}~\bibnamefont{Bloch}},
  \bibinfo{author}{\bibfnamefont{D.~M.} \bibnamefont{Edwards}},
  \bibinfo{author}{\bibfnamefont{M.}~\bibnamefont{Shimizu}}, \bibnamefont{and}
  \bibinfo{author}{\bibfnamefont{J.}~\bibnamefont{Voiron}},
  \bibinfo{journal}{Journal of Physics F: Metal Physics}
  \textbf{\bibinfo{volume}{5}}, \bibinfo{pages}{1217} (\bibinfo{year}{1975}).

\bibitem[{\citenamefont{Takanashi and Shimizu}(1965)}]{TS65}
\bibinfo{author}{\bibfnamefont{T.}~\bibnamefont{Takanashi}} \bibnamefont{and}
  \bibinfo{author}{\bibfnamefont{M.}~\bibnamefont{Shimizu}},
  \bibinfo{journal}{J. Phys. Soc. Japan} \textbf{\bibinfo{volume}{20}},
  \bibinfo{pages}{26} (\bibinfo{year}{1965}).

\bibitem[{\citenamefont{Shimizu and Kato}(1968)}]{SK68}
\bibinfo{author}{\bibfnamefont{M.}~\bibnamefont{Shimizu}} \bibnamefont{and}
  \bibinfo{author}{\bibfnamefont{T.}~\bibnamefont{Kato}},
  \bibinfo{journal}{Phys. Lett.} \textbf{\bibinfo{volume}{27A}},
  \bibinfo{pages}{166} (\bibinfo{year}{1968}).

\bibitem[{\citenamefont{Yeh et~al.}(1975)\citenamefont{Yeh, Chen, Tseng, and
  Fang}}]{YCTF75}
\bibinfo{author}{\bibfnamefont{B.-H.} \bibnamefont{Yeh}},
  \bibinfo{author}{\bibfnamefont{J.}~\bibnamefont{Chen}},
  \bibinfo{author}{\bibfnamefont{P.~K.} \bibnamefont{Tseng}}, \bibnamefont{and}
  \bibinfo{author}{\bibfnamefont{S.-H.} \bibnamefont{Fang}},
  \bibinfo{journal}{Chinese J. Phys} \textbf{\bibinfo{volume}{13}},
  \bibinfo{pages}{1} (\bibinfo{year}{1975}).

\bibitem[{\citenamefont{\ifmmode~\check{S}\else \v{S}\fi{}ipr
  et~al.}(2008)\citenamefont{\ifmmode~\check{S}\else \v{S}\fi{}ipr, Min\'ar,
  Mankovsky, and Ebert}}]{SMME08}
\bibinfo{author}{\bibfnamefont{O.}~\bibnamefont{\ifmmode~\check{S}\else
  \v{S}\fi{}ipr}}, \bibinfo{author}{\bibfnamefont{J.}~\bibnamefont{Min\'ar}},
  \bibinfo{author}{\bibfnamefont{S.}~\bibnamefont{Mankovsky}},
  \bibnamefont{and} \bibinfo{author}{\bibfnamefont{H.}~\bibnamefont{Ebert}},
  \bibinfo{journal}{Phys. Rev. B} \textbf{\bibinfo{volume}{78}},
  \bibinfo{pages}{144403} (\bibinfo{year}{2008}),
  \urlprefix\url{http://link.aps.org/doi/10.1103/PhysRevB.78.144403}.

\bibitem[{\citenamefont{Dahmani}(1985)}]{Dah85}
\bibinfo{author}{\bibfnamefont{C.~E.} \bibnamefont{Dahmani}}, Ph.D. thesis,
  \bibinfo{school}{Louis Pasteur University, Strasbourg}
  (\bibinfo{year}{1985}).

\bibitem[{\citenamefont{Cadeville et~al.}(1986)\citenamefont{Cadeville,
  Dahmani, and Kern}}]{CDK86}
\bibinfo{author}{\bibfnamefont{M.~C.} \bibnamefont{Cadeville}},
  \bibinfo{author}{\bibfnamefont{C.~E.} \bibnamefont{Dahmani}},
  \bibnamefont{and} \bibinfo{author}{\bibfnamefont{F.}~\bibnamefont{Kern}},
  \bibinfo{journal}{J. Magn. Magn. Materials} \textbf{\bibinfo{volume}{55-57}},
  \bibinfo{pages}{1055} (\bibinfo{year}{1986}).

\bibitem[{\citenamefont{Qi et~al.}(1994)\citenamefont{Qi, Skomski, and
  Coey}}]{QSC94}
\bibinfo{author}{\bibfnamefont{Q.}~\bibnamefont{Qi}},
  \bibinfo{author}{\bibfnamefont{R.}~\bibnamefont{Skomski}}, \bibnamefont{and}
  \bibinfo{author}{\bibfnamefont{J.~M.~D.} \bibnamefont{Coey}},
  \bibinfo{journal}{J. Phys.: Condens. Matter} \textbf{\bibinfo{volume}{6}},
  \bibinfo{pages}{3245} (\bibinfo{year}{1994}).

\bibitem[{\citenamefont{Ghosh et~al.}(2001)\citenamefont{Ghosh, Chaudhuri,
  Sanyal, and Mookerjee}}]{GCS01}
\bibinfo{author}{\bibfnamefont{S.}~\bibnamefont{Ghosh}},
  \bibinfo{author}{\bibfnamefont{C.~B.} \bibnamefont{Chaudhuri}},
  \bibinfo{author}{\bibfnamefont{B.}~\bibnamefont{Sanyal}}, \bibnamefont{and}
  \bibinfo{author}{\bibfnamefont{A.}~\bibnamefont{Mookerjee}},
  \bibinfo{journal}{J. Magn. Magn. Materials} \textbf{\bibinfo{volume}{234}},
  \bibinfo{pages}{100} (\bibinfo{year}{2001}).

\bibitem[{\citenamefont{Mohn and Wohlfarth}(1987)}]{MW87}
\bibinfo{author}{\bibfnamefont{P.}~\bibnamefont{Mohn}} \bibnamefont{and}
  \bibinfo{author}{\bibfnamefont{E.~P.} \bibnamefont{Wohlfarth}},
  \bibinfo{journal}{J. Phys. F: Met. Phys.} \textbf{\bibinfo{volume}{17}},
  \bibinfo{pages}{2421} (\bibinfo{year}{1987}).

\bibitem[{\citenamefont{Deng et~al.}(2001)\citenamefont{Deng, Freyer,
  Voitl\"ander, and Ebert}}]{DFVE01}
\bibinfo{author}{\bibfnamefont{M.}~\bibnamefont{Deng}},
  \bibinfo{author}{\bibfnamefont{H.}~\bibnamefont{Freyer}},
  \bibinfo{author}{\bibfnamefont{J.}~\bibnamefont{Voitl\"ander}},
  \bibnamefont{and} \bibinfo{author}{\bibfnamefont{H.}~\bibnamefont{Ebert}},
  \bibinfo{journal}{J. Phys.: Cond. Mat.} \textbf{\bibinfo{volume}{13}},
  \bibinfo{pages}{8551} (\bibinfo{year}{2001}).

\bibitem[{\citenamefont{Rose}(1961)}]{Ros61}
\bibinfo{author}{\bibfnamefont{M.~E.} \bibnamefont{Rose}},
  \emph{\bibinfo{title}{Relativistic Electron Theory}}
  (\bibinfo{publisher}{Wiley}, \bibinfo{address}{New York},
  \bibinfo{year}{1961}).

\bibitem[{\citenamefont{Mankovsky and Ebert}(2006)}]{ME06}
\bibinfo{author}{\bibfnamefont{S.}~\bibnamefont{Mankovsky}} \bibnamefont{and}
  \bibinfo{author}{\bibfnamefont{H.}~\bibnamefont{Ebert}},
  \bibinfo{journal}{Phys. Rev. B} \textbf{\bibinfo{volume}{74}},
  \bibinfo{pages}{54414} (\bibinfo{year}{2006}).

\end{thebibliography}

\end{document}